\documentclass[12pt]{article}

\usepackage{scicite}

\usepackage{times}

\newenvironment{sciabstract}{%
\begin{quote} \bf}
{\end{quote}}

\usepackage{graphicx}
\usepackage[per-mode = symbol]{siunitx}
\DeclareSIUnit{\BL}{BL}
\DeclareSIUnit{\SLPM}{SLPM}

\usepackage{enumitem}

\usepackage{listings}
\usepackage{xcolor}
\definecolor{codegreen}{rgb}{0,0.6,0}
\definecolor{codegray}{rgb}{0.5,0.5,0.5}
\definecolor{codepurple}{rgb}{0.58,0,0.82}
\definecolor{backcolour}{rgb}{0.95,0.95,0.92}
\lstdefinestyle{mystyle}{
    backgroundcolor=\color{backcolour},   
    commentstyle=\color{codegreen},
    keywordstyle=\color{magenta},
    numberstyle=\tiny\color{codegray},
    stringstyle=\color{codepurple},
    basicstyle=\ttfamily\footnotesize,
    breakatwhitespace=false,         
    breaklines=true,                 
    captionpos=b,                    
    keepspaces=true,                 
    numbers=left,                    
    numbersep=5pt,                  
    showspaces=false,                
    showstringspaces=false,
    showtabs=false,                  
    tabsize=2
}
\usepackage{physunits}

\usepackage{ulem} 
\usepackage{amsthm}

\usepackage{amsmath}
\usepackage{comment}
\usepackage{soul}
\usepackage{hyperref}
\usepackage{wrapfig}

\usepackage[pages=all, placement=outer, color=black, scale=1, angle=90, vshift=100mm, hshift=0mm]{background}

\backgroundsetup{
  contents={\textsf{\large Author Manuscript. Definitive version published in \emph{Science} on May 8th, 2025; DOI: 10.1126/science.adr3661}}
}

\title{Physical synchronization of soft self-oscillating limbs for fast and autonomous locomotion\\ \small{\;} \\ \large{Accepted Preprint. Published in edited form as: \emph{Science}, May 8th, 2025  \href{https://www.science.org/doi/10.1126/science.adr3661}{doi.org/10.1126/science.adr3661}
}}

\author
{Alberto Comoretto$^{1}$, Harmannus A.H. Schomaker,$^{1}$ Johannes T.B. Overvelde$^{1,\ast}$\\
\\
\normalsize{$^{1}$Autonomous Matter Department, AMOLF, Science Park 104, 1098 XG Amsterdam, The Netherlands}
\\
\normalsize{$^\ast$To whom correspondence should be addressed; E-mail:  overvelde@amolf.nl.}
}

\date{}

\begin{document}


\baselineskip24pt


\maketitle


\begin{sciabstract}
Animals achieve robust locomotion by offloading regulation from the brain to physical couplings within the body. In contrast, locomotion in artificial systems often depends on centralized processors. We introduce a rapid and autonomous locomotion strategy with synchronized gaits emerging through physical interactions between self-oscillating limbs and the environment, without control signals. Each limb is a single soft tube that only requires constant flow of air to perform cyclic stepping motions at frequencies reaching 300 hertz. By combining several of these self-oscillating limbs, their physical synchronization enables locomotion speeds that are orders of magnitude faster than comparable state-of-the-art. Through body-environment dynamics, these seemingly simple devices exhibit autonomy, including obstacle avoidance, amphibious gait transitions, and phototaxis.
\\ \small{\;}

\textnormal{This is the author’s version of the work. The definitive version was published in \emph{Science} on May 8th, 2025; DOI: \href{https://www.science.org/doi/10.1126/science.adr3661}{10.1126/science.adr3661}. This author manuscript is subject to AAAS Science's License to publish. \\\raisebox{-0.8em}{\includegraphics[width=0.1\textwidth]{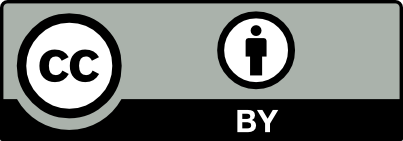}} \; This work is licensed under CC BY 4.0. \href{https://creativecommons.org/licenses/by/4.0}{https://creativecommons.org/licenses/by/4.0}.}

\end{sciabstract}


\newpage

Nature masters the complex problem of locomotion through embodied solutions, harnessing the synergy between the nervous system, body, and environment \cite{brain_has_body}. The foundation of animal locomotion lies in the periodic and asymmetric motion of individual limbs \cite{biorobotics_agile_locomotion,how_animals_move} (Fig.~1A). Multiple oscillating limbs are typically coordinated through a variety of embodied couplings, which are computationally and metabolically inexpensive by diminishing or even eliminating the need for sequential individual inputs from a central brain \cite{how_animals_move}. These embodied couplings are often implemented through both internal neural connections (i.e., explicitly) and external interactions with the environment (i.e., implicitly). For instance, stick insects achieve synchronized walking gaits through both explicit neural connections and implicit body-environment interactions, avoiding centralized patterning \cite{insect_walking_coupling}. Sea stars explicitly coordinate their five arms in a decentralized fashion through a nerve ring \cite{seastar_nerve_ring} and occasionally exhibit a fast bouncing gait as an escape response, with their hundreds of tube feet achieving synchronization through implicit mechanical coupling with the external substrate \cite{seastar_heydari} (Fig.~1A).

Inspired by nature, robots can delegate the locomotion task to their bodies \cite{embodiment_robotics, leon}, thus minimizing energy and time costs associated with a central computer. For instance, by harnessing the dynamics of the body, rigid robots based on passive-dynamic walkers \cite{passive_dynamic_walkers} and synergy-based quadrupeds \cite{stella_paws} reduce, but do not eliminate, the amount of control required for locomotion. Without processors, soft robots based on twisting liquid crystal elastomers \cite{twisting_LCE} and elasto-active structures \cite{bucklebots} harness shape reconfiguration to avoid obstacles autonomously. However, due to the lack of limbs, their applicability is limited to specific tasks and environments \cite{legged_robots_book}, in contrast to the wide spectrum of robust behaviors typical of animals, which is often enabled by interactions between multiple self-oscillators \cite{salamander_robot}.

Among soft-limbed robots, fluidic circuits \cite{hardware_methods_fluidic_control} sequence the activation of limbs \cite{octobot} for walking gaits \cite{gorissen_sequencing, drotman_turtle, luuk, 3D_printed_tauber} without the need for electronic processors. However, fluidic circuits still emulate their electronic counterpart \cite{digital_logic_soft}, involving multiple, macroscopic, digital or analog components that deliver sequential control signals. This architecture leads to energy losses and delay across the fluidic network, causing slow sequencing of the limbs in the order of one hertz, with consequent ineffective locomotion of only a few body lengths (BLs) per minute \cite{harnessing_viscous_flow,luuk,flat_ring_tube_valve,gorissen_sequencing,drotman_turtle,soft_bistable_valve,bro,3D_printed_tauber,cheap_valves}, which is impractical for most real-world applications \cite{3D_soft_radiation,robots_environmental_monitoring,search_and_rescue}. Moreover, autonomous behavior of walkers with fluidic processors remains elusive, with the exception of one-time-use touch sensing \cite{drotman_turtle} and reprogrammable sequencing of non-integrated soft fingers \cite{luuk}.

Inspired by the movement principles of animals that do not require centralized processing, we aim to harness physical synchronization of limbs for rapid and autonomous locomotion in soft-limbed robots with embedded fluidic circuits. To do so, in this work we introduce three levels of behavioral hierarchy (Fig.~1B) based on \emph{i)} asymmetric self-oscillating motion at the limb level, \emph{ii)} explicit internal fluidic coupling leading to synchronized gaits, and \emph{iii)} implicit coupling between body dynamics and environmental interactions resulting in autonomous behavior. Based on these principles, we develop robotic demonstrations that display rapid locomotion and adaptive behaviors, providing fundamental and general insights into how to instill autonomy in systems without electronic, electronic-like mechanical \cite{mechanical_computing}, or fluidic \cite{digital_logic_soft} processors.

\paragraph*{Self-oscillating limb}

To enable rapid locomotion without a central processor, here we develop a limb that undergoes periodic and asymmetric motions by harnessing a self-oscillating behavior of thin soft tubes, analogous to oscillations occurring in flat tubes with flowing water \cite{flat_ring_tube}, and reminiscent of promotional air dancers \cite{tubeguy_patent} often seen by the roadside. We build the limb by simply bending a thin-walled commercial silicone tube $\SI{180}{\degree}$ and constraining it at the inlet and outlet in a 3D-printed holder (Fig.~1C and fig.~S1). If no airflow is provided, the tube displays two stable states with either one \cite{kink_valves} or two kinks (Fig.~1C,D). However, when we apply a constant airflow of $15$ standard liter per minute (\SI{}{\SLPM}) to the inlet on the left side of the tube, the tube starts to spontaneously oscillate between states with one and two kinks, at a frequency of approximately \SI{100}{\hertz} (Fig.~1E and movie~S1) (standard units of airflow in table~S1).

\begin{figure}[t!]
\centering
\includegraphics[width=12cm]{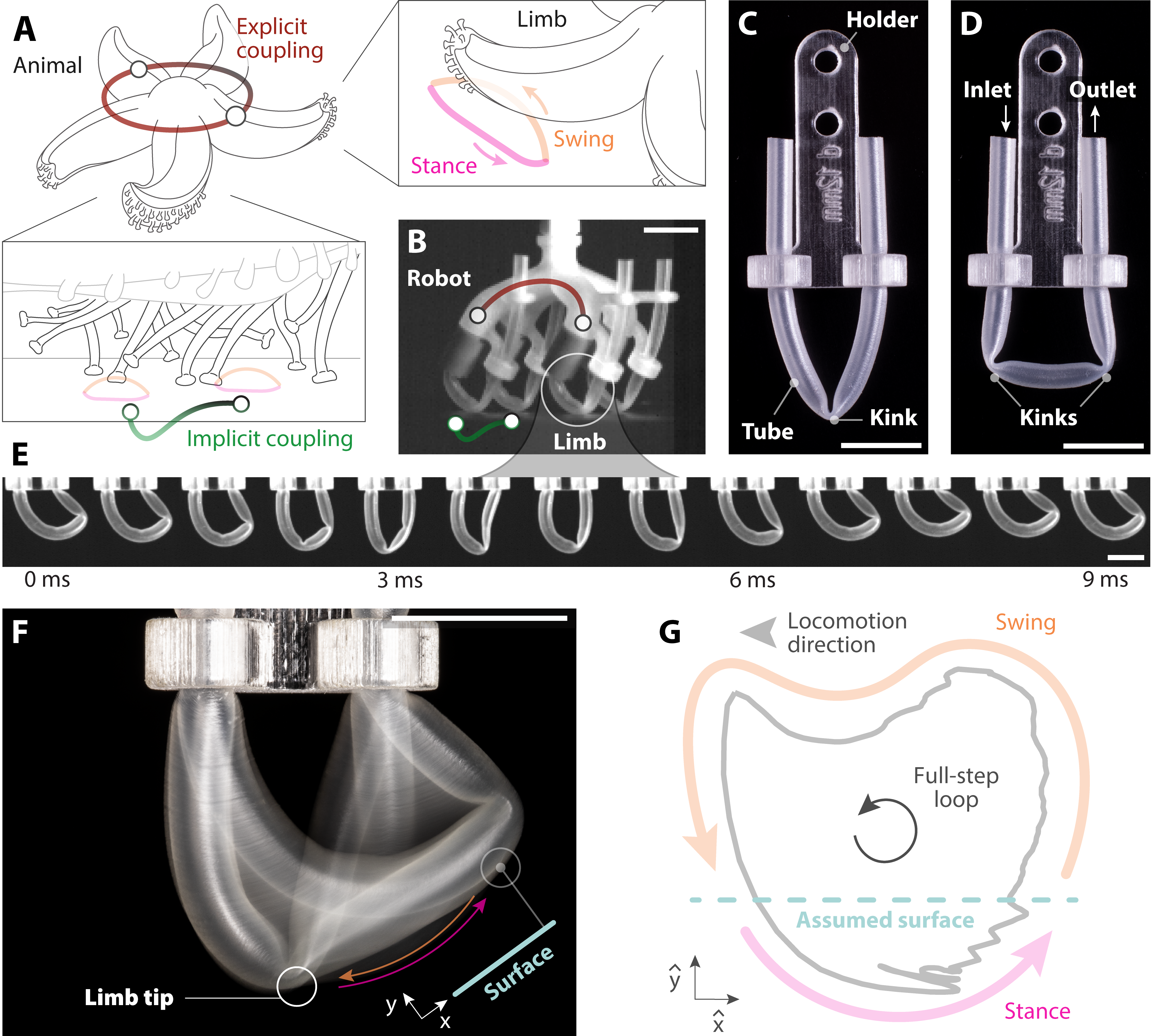}
\par\medskip
    \caption{\label{fig1} \textbf{Self-oscillating limbs that cyclically perform full-step motions.} (\textbf{A}) Animals locomote by coordinating multiple limbs via explicit coupling through neural connections or implicit coupling through interaction with the environment \cite{insect_walking_coupling,seastar_heydari,seastar_nerve_ring}. Each limb of the animal performs oscillating and asymmetric (full-step) motions with stance and swing phases \cite{biorobotics_agile_locomotion,how_animals_move}. (\textbf{B}) We exploit these principles of explicit and implicit couplings between self-oscillating limbs for autonomous locomotion in robots. Our artificial limb is a soft tube bent 180° that, in static conditions, displays (\textbf{C}) one or (\textbf{D}) two kinks. (\textbf{E}) When constant airflow of $15$ standard liter per minute (\SI{}{\SLPM}) is provided at the inlet on the left end, the tube self-oscillates at a frequency of \SI{115}{\hertz} (snapshots of one oscillation cycle). (\textbf{F}) The tip of the limb is the point on the tube closest to a defined surface (photograph with \SI{0.5}{\second} exposure time, capturing $\sim50$ consecutive oscillations). (\textbf{G}) The tip cyclically undergoes a full-step loop trajectory, with a stance phase followed by a swing phase (the reported tip trajectory coordinates $\hat{x}$ and $\hat{y}$ are normalized). Scale bars are \SI{1}{\centi\meter}.
    }
\end{figure}

The motion of the tube during the oscillation cycle is intrinsically asymmetric due to the applied directed airflow. As a result, the tip of the limb, which we define as the point on the tube closest to a defined surface (Fig.~1F and fig.~S10), traces a hysteretic trajectory in the $x$-$y$ plane orthogonal to the surface (Fig.~1G and fig.~S10). The tube, therefore, acts as a limb that undergoes a full-step motion, with a periodic closed-loop sequence of stance and swing phases (Fig.~1L and movie~S1), reminiscent of animals' limbs \cite{how_animals_move} (Fig.~1A). For constant input airflow, we delegate both the oscillation generation and the sequencing of the tip motion within each cycle directly to the limb itself, bypassing the requirement of additional circuitry \cite{hardware_methods_fluidic_control}, either fluidic \cite{drotman_turtle,3D_printed_tauber} or electronic \cite{untethered_multigait}.

We next aim to empirically understand the switching behavior between one and two kinks in the self-oscillating limb. To do so, we detect the outer and inner edges of the tube (Fig.~2A and Supplementary Material section~M3) and we map them on a new coordinate system along the center line of the tube (Fig.~2B, fig.~S5 and fig.~S12). Focusing on the local minima of the tube width, we indeed identify one or two kinks (Fig.~2B and movie~S2). We refer to the kink with the smallest width as the dominant kink, and the second kink (if present at that instant) as the non-dominant kink.

From the location of the kinks in time (Fig.~2C and fig.~S12), and the inlet pressure (Fig.~2D), we can identify four key steps during a single oscillation cycle. At $\sim\SI{5}{\milli\second}$ (Fig.~2E), we observe two kinks staying approximately at the same location on the tube until $\sim\SI{35}{\milli\second}$ (Fig.~2F). During this time, the pressure at the inlet decreases, causing the non-dominant kink to increasingly sharpen (fig.~S21). Between $\sim\SI{35}{\milli\second}$ and $\sim\SI{40}{\milli\second}$ the non-dominant kink becomes dominant, while the other kink disappears at $\sim\SI{40}{\milli\second}$ (Fig.~2G). Until $\sim\SI{60}{\milli\second}$, the only kink in the tube starts to travel along the tube (Fig.~2H), as a result of an increase in pressure before the kink (fig.~S14). Finally, after $\sim\SI{60}{\milli\second}$, a new kink forms upstream again due to an increase in bending moment (fig.~S21). This state (equivalent to Figure~2E) is characterized by a lower fluidic resistance than the single kink state in Figure~2G, as both kinks have a larger angle (fig.~S21), and thus causes pressure to drop again. 

\begin{figure}[t!]
\centering
\includegraphics[width=12cm]{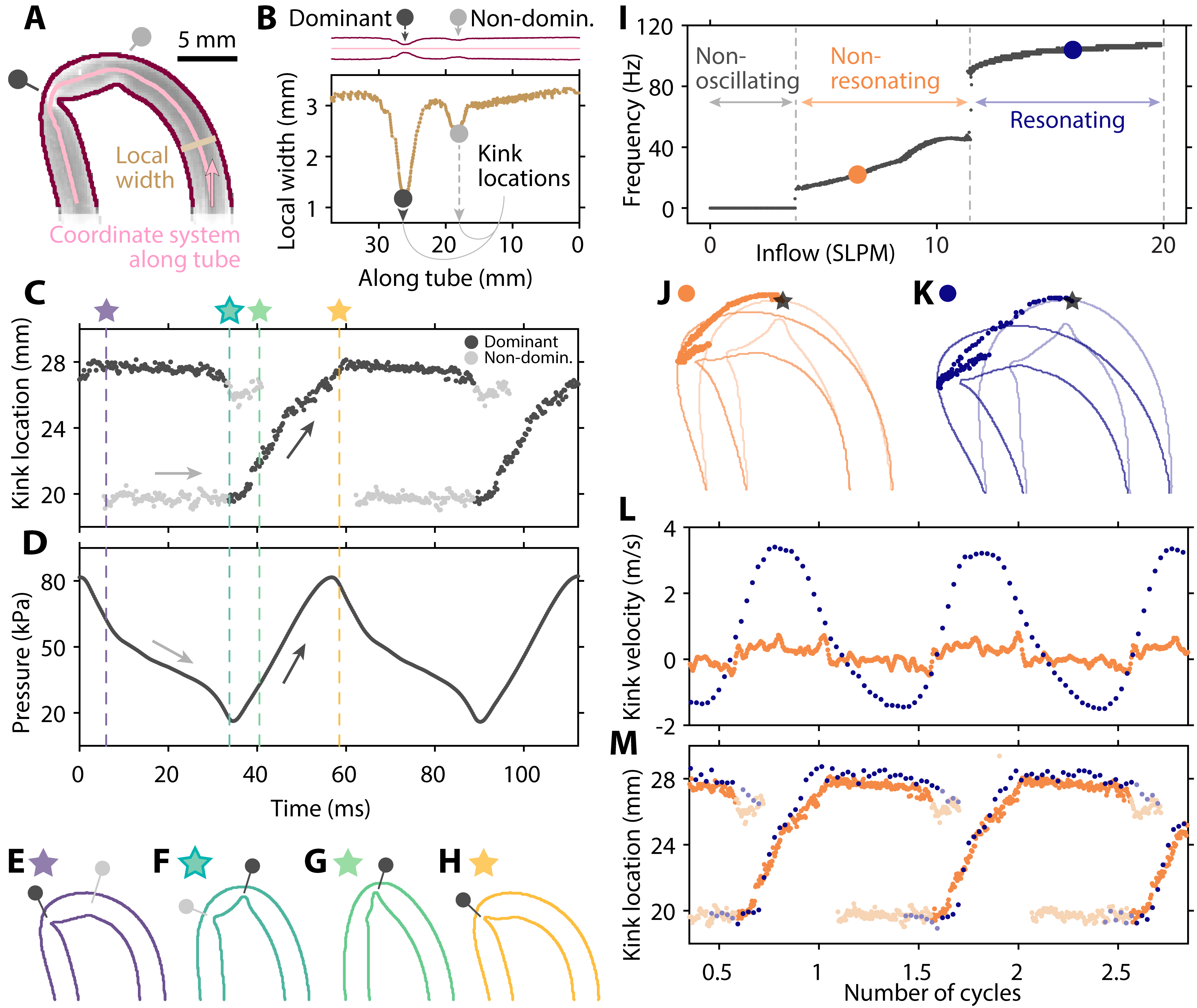}
\par\medskip
\caption{\label{fig2} \textbf{Interplay between pressure and kinks' state enables self-oscillation at a range of frequencies.} (\textbf{A}) Detected edges of the tube (dark red) and coordinate system along the tube (pink). (\textbf{B}) The dominant and non-dominant kinks correspond to the local minima (black and grey dots) of the local width along the tube. The  (\textbf{C}) location of the kinks along the tube and (\textbf{D}) pressure inside the tube are coupled. (\textbf{E}, \textbf{F}, \textbf{G}, \textbf{H}) State of the tube at four instants of the oscillating cycle. (\textbf{I}) The oscillation frequency displays three regimes for different inflow rates. Orange and blue dots correspond to the non-resonating and resonating study cases at $6.5$ \SI{}{\SLPM} and $16$ \SI{}{\SLPM}, respectively. The tube has lower structural displacement in the (\textbf{J}) non-resonating domain than in the (\textbf{K}) resonating domain, as shown by the kink covering a shorter distance. (\textbf{L}) In the resonating case, the structure undergoes high quasi-sinusoidal velocities, in comparison to the near-zero velocities of the non-resonating case. (\textbf{M}) The kink locations along the tube itself, for the two cases, overlap.}
\end{figure}

From a basic mass-spring model (Supplementary Material section~S1), we gain insight into three ingredients that lead to the self-oscillating behavior (fig.~S6): \emph{i)} a local nonlinear torque-angle curve under bending leads to localized kinking of the tube (fig.~S8); \emph{ii)} the local flow resistance increases at the kink location, leading to increased pressure before the kink (fig.~S20); \emph{iii)} pressurization of the tube increases the torque required for kinking and unkinking (fig.~S8 and fig.~S9).  Overall, for a constant inflow of air, the internal pressure varies depending on the kinks' resistance. In turn, pressure decrease favors kink growth, and pressure increase causes kinks to travel. This interaction between tube stiffening, kink growth and travel, and change in resistance sets up a hysteric actuation loop (fig.~S14) that results in the self-oscillating behavior for constant inflow. 

We find that the frequency of the self-oscillation can be tuned by varying the flowrate through the tube. When we sweep the inflow between $0$ and $20$ \SI{}{\SLPM}, we observe three domains, with sharp transitions between them (Fig.~2I). For inflow values below $\sim3.8$ \SI{}{\SLPM}, the tube leaks and does not oscillate (fig.~S13). For flowrates between $\sim3.8$ \SI{}{\SLPM} and $\sim11.4$ \SI{}{\SLPM}, the system oscillates at frequencies between $\sim13$ and $\sim\SI{45}{\hertz}$. For even higher flowrates the frequency suddenly jumps to higher frequencies above $\sim\SI{90}{\hertz}$.

This last sudden jump in frequency seems to be the result of structural resonance (movie~S2). This can be seen from the increase in the distance traveled by the kink (Fig.~2J,K and fig.~S12) and by the dramatic increase in the kink velocity, that approaches a sinusoidal trend (Fig.~2L) (Supplementary Material section~M3 for the definitions of kink distance and velocity). In comparison, the kink location along the tube does not change considerably when the flow is increased (Fig.~2M), therefore not constituting the origin of the sudden increase in frequency. We conclude that we can significantly increase the oscillation frequency by exploiting the resonance of the structure, as well as by varying design parameters (fig.~S11 and fig.~18).

\paragraph*{Explicit internal coupling of multiple limbs for ultrafast locomotion}

The individual self-oscillating limb requires integration with other limbs in a multi-limbed system to enable locomotion and autonomy in robotic applications. With the goal of synchronizing the activation of several limbs to generate specific gaits, we couple two limbs by connecting them in parallel to a single input, using two identical coupling tubes with an inner diameter of \SI{2}{\milli\meter} (Fig.~3A). When providing a constant airflow to the inlet we observe that coupling tubes shorter than $\SI{12}{\centi\meter}$ lead to in-phase synchronization. The result is that the two kinks travel simultaneously (Fig.~3B) and the pressure signals align (Fig.~3C and movie~S3), even though the natural frequencies of the two limbs are not identical and differ by $\sim\SI{5}{\hertz}$ ($\sim2\%$) (fig.~S15). In contrast, coupling tubes longer than $\SI{12}{\centi\meter}$ result in anti-phase synchronization where the limbs alternatively activate with a phase shift of $\sim\SI{180}{\degree}$ (Fig.~3D,E and movie~S3).

\begin{figure}[t!]
\centering
\includegraphics[width=18.4cm]{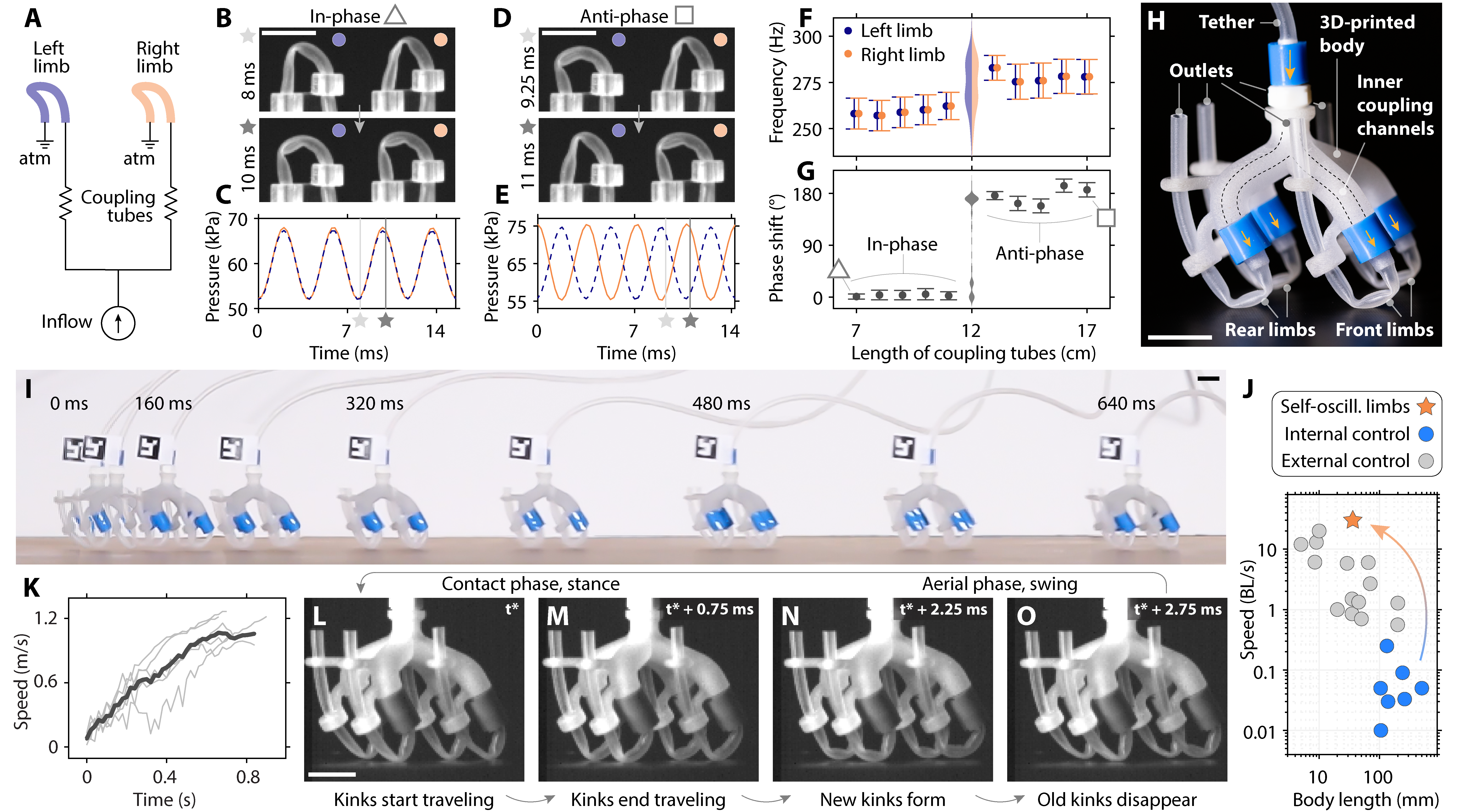}
\par\medskip
\caption{\label{fig3} \textbf{Synchronization of multiple limbs through explicit, internal coupling for ultrafast locomotion.} (\textbf{A}) We couple two limbs in parallel to the same input flow source of $15$ \SI{}{\SLPM} with two identical silicone tubes. We observe in-phase synchronization, with (\textbf{B}) simultaneous kink traveling and (\textbf{C}) aligned pressure signals, or anti-phase synchronization, with (\textbf{D}, \textbf{E}) alternate activation of the limbs. We scan the length of the coupling tubes, observing two separated in-phase and anti-phase domains, as (\textbf{F}) the oscillation frequency of left and right limbs match, and (\textbf{G}) the phase-shift is either $\sim\SI{0}{\degree}$ or $\sim\SI{180}{\degree}$. (\textbf{H}) The tethered robot has four limbs connected to a 3D-printed monolithic body, with four inner coupling channels. (\textbf{I}) The robot achieves ultrafast locomotion on a flat surface (speed $\sim\SI{30}{\BL\per\second}$, \SI{1.1}{\meter\per\second}). (\textbf{J}) Comparison between tethered soft robots with internal and external control and our robot equipped with synchronizing self-oscillating limbs, in terms of relative speed and body length. (\textbf{K}) Speed of the robot for six runs (grey) and mean speed (black). The four limbs, within $\sim\SI{3}{\milli\second}$, simultaneously go through a (\textbf{L}, \textbf{M}) stance phase, followed by a (\textbf{N}, \textbf{O}) swing phase. All scale bars are \SI{1}{\centi\meter}.}
\end{figure}

When scanning a wide range of the coupling tubes' length, we note a sharp transition between the in-phase and anti-phase eigenmodes (Fig.~3F,G). When placed at this transition, the system continuously switches between the in-phase and anti-phase eigenmodes (violin plot at \SI{12}{\centi\meter} in Figure~3G, figure~S15 and movie~S3).

This synchronization effect is reminiscent of strong coupling between the two oscillators \cite{synchronization_universal_concept}. In soft systems, strong coupling has previously been observed in mechanically-coupled liquid crystalline oscillators \cite{coupled_liquid_crystal_oscill}, while here the coupling is induced by the fluidic interconnections.

Based on these findings, we build a robot with four strongly-coupled self-oscillating limbs, assembled onto a 3D-printed body with short inner channels with length $\sim\SI{1.5}{\centi\meter}$, connected in parallel to a single tether (Fig.~3H, fig.~S3 and fig.~S4). We orient the limbs with a \SI{30}{\degree} angle to the surface to optimize the effective stance (fig.~S10). When we provide a constant inflow of $\sim 28$ \SI{}{\SLPM} to the tether (fig.~S22), the robot accelerates (Fig.~3I and movie~S4), reaching a steady-state speed of $30\pm\SI{2.5}{\BL\per\second}$ (\SI{1.1}{\meter\per\second}) (Fig.~3J,K), with a response time of \SI{0.66}{\second} (Fig.~3K and fig.~S22) and Froude number $\sim20.6$ (Supplementary Material section~S4).  This speed is two orders of magnitude higher than comparable state-of-the-art tethered robots with internal actuation sequencing \cite{3D_printed_tauber}, and similar to current ultrafast tethered soft robots that need external control \cite{insect_ultrafast} (Fig.~3J, fig.~S27 and table~S2).

Looking closely at the gait in Figure~3L-O and movie~S4, we observe that all the limbs autonomously activate in synchrony, since we used short inner coupling channels. The robot runs with a stotting gait, typical of gazelles \cite{pronking}, reached after some initial transient asynchronous behavior which lasts $\sim\SI{0.2}{\second}$ (Fig.~3K and movie~S4). Note that the four limbs oscillate at a frequency of $\sim\SI{300}{\hertz}$ (Fig.~3L-O), about three times higher than the case of the single limb that we analyzed in Figure~2, because the tubes are smaller in diameter by a factor $\sim0.8$ and shorter in length by a factor $\sim0.5$ (Supplementary Material section~S4).

\paragraph*{Implicit environmental coupling for fast and autonomous locomotion}

Even though our robot achieved ultrafast locomotion without external control, it still required a tether that provides a power of $\sim\SI{85}{\watt}$, leading to a relatively high cost of transport of $\sim1926$ (Supplementary Material section~S4). At the moment, it is not possible to generate this power from a lightweight on-board pressure source, making this robot not directly suited for untethered applications. We find that the main limitation comes from the required flow of $\sim3.8$ \SI{}{\SLPM} for the individual limb to not leak through the kink and start oscillating (Fig.~2I and fig.~S13). We hypothesize that an increase in kink resistance can reduce the leakage and thus the minimum flow needed to achieve self-oscillation, thus reducing the required power. 

We modified the design of our tubes by heat-sealing two thermoplastic polyurethane (TPU) sheets along two parallel lines. We mount this TPU component to a hinge joint to obtain a self-oscillating pouch limb (fig.~S2). This updated limb has a much higher flow resistance upon kinking (fig.~S21) and performs the full-step oscillation (Fig.~4A) with a minimum input airflow of only $0.1$ \SI{}{\SLPM} (fig.~S16). Compared to the silicone tube limb, we observe a lower oscillation frequency of maximum $\sim\SI{3.5}{\hertz}$ (fig.~S16 and fig.~S18). This is likely due to the higher geometric volume of air required for the kink to travel ($\sim\SI{2.5}{\milli\liter}$ compared to $\sim\SI{0.04}{\milli\liter}$, Supplementary Material section~S2) which leads to a longer kink traveling time, and the absence of a resonant mode. However, this frequency reduction is compensated by a larger stroke per cycle of the pouch limb (Fig.~4B, fig.~S16 and movie~S1).

We build an untethered robot with two pouch tubes as soft limbs, each requiring only $\sim\SI{0.06}{\watt}$ of fluidic power (Supplementary Material section~S2 and fig.~S17), and each connected to its own \SI{3}{\volt} air pump, powered by a \SI{3.7}{\volt} LiPo battery with \SI{380}{\milli\ampere\hour} (Fig.~4C, fig.~S3 and fig.~S4). The total weight of the assembled robot equals \SI{76.7}{\gram}, which is $\sim6$ times lower than the maximum force two pouch-based limbs can provide (fig.~S19). When we turn the pumps on, the robot starts to cyclically hop at a rate of $\sim\SI{2}{\hertz}$ (movie~S5), with each hopping cycle characterized by a stance phase followed by a swing phase (Fig.~4D). The untethered robot moves with a speed of $1.93\pm\SI{0.07}{\BL\per\second}$ ($\SI{18.1}{\centi\meter\per\second}$) (fig.~S22), which is one order of magnitude faster than untethered soft fluidic robots \cite{cheap_valves}, and comparable to state-of-the-art untethered soft robots \cite{hair_clip} (fig.~S27 and table~S3). Under the tested conditions, the robot locomotes with a cost of transport $\sim11$, in the same order of land animals of comparable size such as mice \cite{adaptive_morphogenesis}, and with Froude number $\sim0.16$ (Supplementary Material section~S4).

\begin{figure}[t!]
\centering
\includegraphics[width=12.1cm]{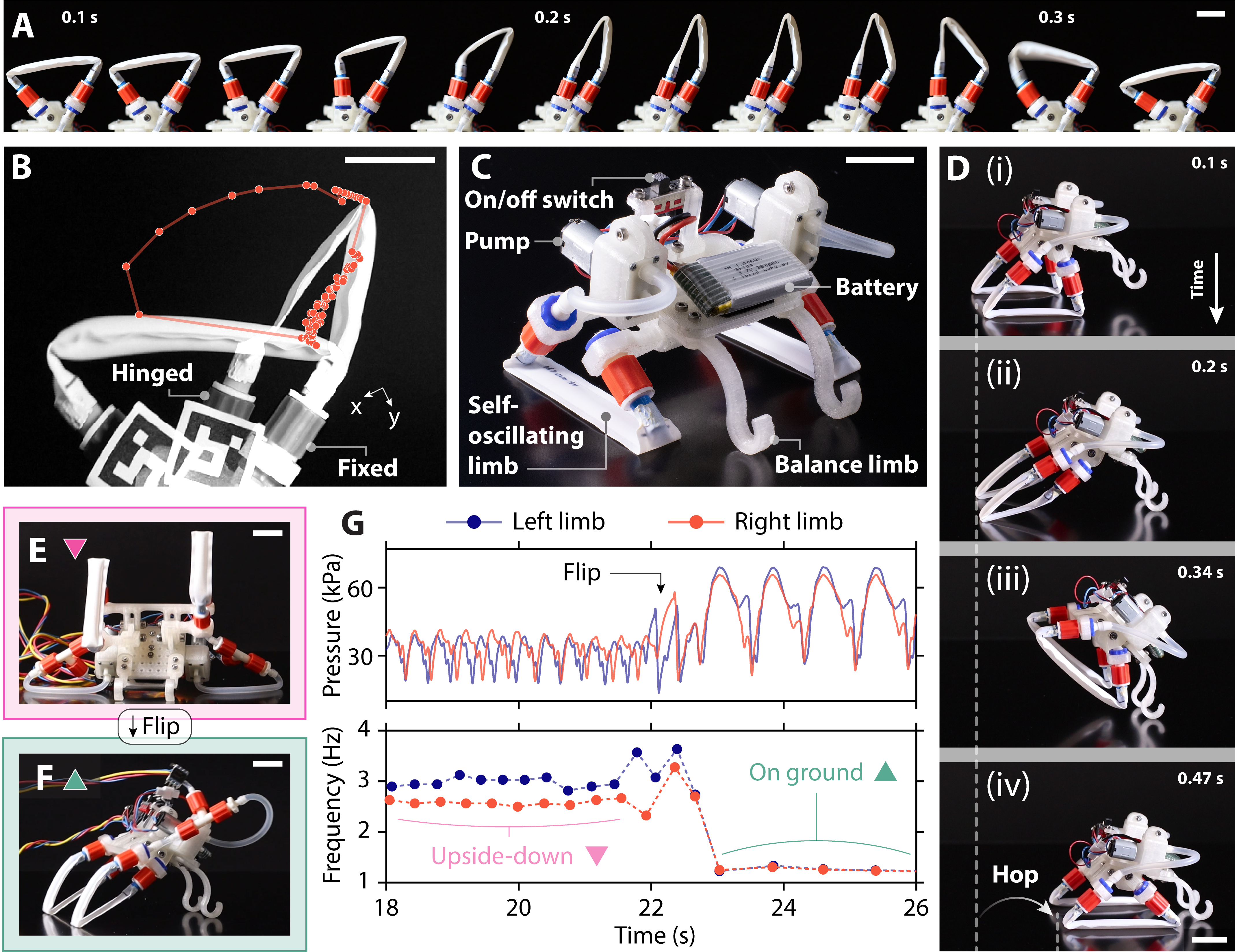}
\par\medskip
\caption{\label{fig4} \textbf{Synchronization of limbs through implicit interaction for fast, untethered locomotion.} The updated pouch limb (\textbf{A}) cyclically performs full-step motions with a low inflow of $0.3$ \SI{}{\SLPM}, (\textbf{B}) displaying large hysteresis and stroke enabled by a hinge joint. (\textbf{C}) We mount two self-oscillating limbs on an untethered robot that carries a LiPo battery and two pumps. (\textbf{D}) The robot cyclically hops with (i, ii) a stance phase followed by (iii, iv) a swing phase. (\textbf{E}) The two limbs are not synchronized when the system is upside down, because they are independently powered by the two pumps. (\textbf{F}) When the robot interacts with the ground, the two limbs synchronize. (\textbf{G}) While interacting, the pressure signals of the two limbs increase, and the frequencies equalize. All scale bars are \SI{2}{\centi\meter}.
}
\end{figure}

In this untethered scenario, the robot's high speed is attributed to the synchronization of the soft limbs. However, in this case, synchronization emerges due to implicit interactions with the environment, and not through embedded fluidic connections as was the case for the tethered robot, as each limb has its own power source. In principle, we observe that the two limbs actuate at different natural frequencies and thus out of phase when placed upside down (Fig.~4E and movie~S5). When we flip the robot into the working position where the two limbs interact with the ground, either on a flat surface (Fig.~4F) or on gravel (fig.~S23 and fig.~24), the limbs start to actuate simultaneously and in-phase, while requiring higher pressures (Fig.~4G  and movie~S5). This in-phase synchronization results from the positive coupling between the limbs, where the activation of one limb stimulates the activation of the other limb. Additional mass stabilizes the in-phase synchronized mode and improves the tolerance to imbalance in left and right input flows (Supplementary Material section~S4 and fig.~S26).

Harnessing such implicit interactions to achieve synchronization also enables autonomous behaviors, as external cues can affect the coupling between the body dynamics and the environment. For instance, through implicit coupling with its surroundings, the robot autonomously transitions to a new locomotion gait when interacting with an aquatic environment (Fig.~5A and movie~S6), without requiring any control input \cite{salamander_robot} or morphological change \cite{adaptive_morphogenesis}. After diving into the water, the in-phase hopping gait within seconds transitions to an anti-phase gait (Fig.~5B), which is stable in steady-state conditions, with a phase shift of $181\pm\SI{8}{\degree}$ (Supplementary Material section S4). The anti-phase gait seems to be stabilized by a sideways swaying of the robot, leading to negative coupling in which the activation of a limb suppresses the activation of the other limb (movie~S6). To confirm this, we check that manually constraining the swaying motion causes the limbs to not synchronize anymore (movie~S6). Therefore, the coupling between the body dynamics and the surrounding medium determines the stability of a specific gait for the robot and, thus, its behavior.

When encountering an obstacle, too, the mechanical interactions between the body and an obstacle cause temporary asynchronous activation of the legs, resulting in the robot steering in place and avoiding the obstacle (Fig.~5C and movie~S6). By harnessing these environmental cues, the robot changes behavior and autonomously escapes a U-shaped obstruction after noisy and consecutive random changes of direction (Fig.~S25 and movie~S6).

\begin{figure}[t!]
\centering
\includegraphics[width=12.1cm]{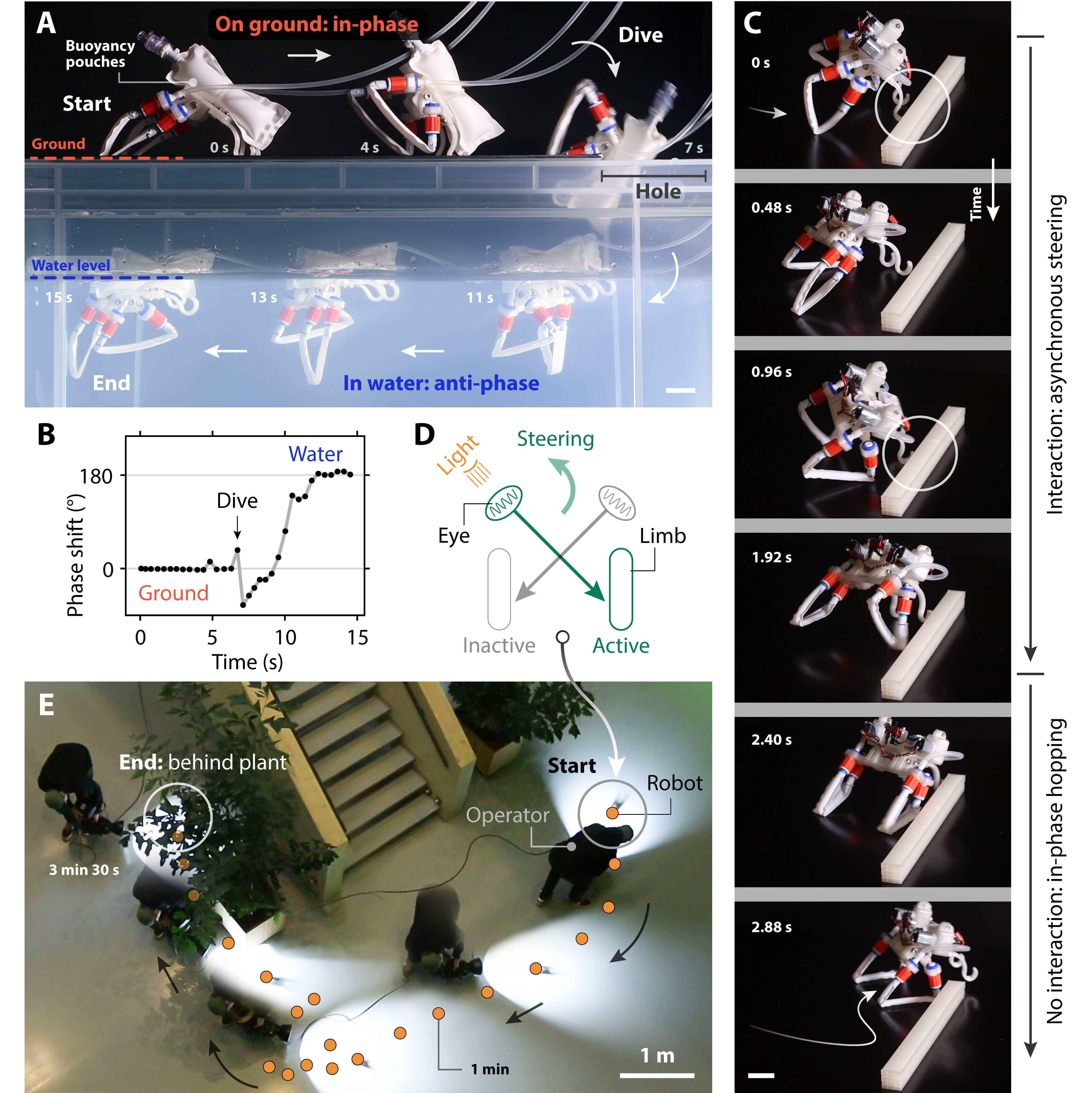}
\par\medskip
\caption{\label{fig5} \textbf{Autonomy through physical interactions with the environment.} (\textbf{A}) After diving into the water, the robot equipped with buoyancy pouches ($\SI{50}{\milli\liter}$ of air) autonomously transitions to an anti-phase swimming gait, through the implicit coupling with the new aquatic environment. (\textbf{B}) The phase shift between the two self-oscillating limbs is $\sim\SI{0}{\degree}$ (in-phase) when hopping on the ground and spontaneously transitions to $\sim\SI{180}{\degree}$ (anti-phase) when interacting with water. (\textbf{C}) When encountering an obstacle, the mechanical interactions (white circles) cause the limbs to activate asynchronously and, as a consequence, the robot steers in place, avoiding the obstacle. (\textbf{D}) To provide the robot with a high-level sense of direction, and inspired by Braitenberg's `aggressive vehicle' \cite{vehicles}, we cross-link light sensors (eyes) and the pumps so that a limb is active when the opposite eye detects light. (\textbf{E}) This robot achieves autonomous phototaxis by steering in place when only one eye is active and hopping forward when both eyes are, following an operator that carries a light in a real-world environment. Wherever not otherwise stated, scale bars are \SI{2}{\centi\meter}.}
\end{figure}

As the amphibious locomotion is unidirectional and the body-obstacle interactions result in random outcomes (movie~S6), we propose that additional sensing is still needed to provide the robot with a high-level sense of direction. We take inspiration from Braitenberg's `aggressive vehicle 2b' \cite{vehicles}, and design internal connections by equipping the robot with two `eyes' (LDR light sensors) that activate the opposite soft limbs when sensing light (Fig.~5D and fig.~S4). This relatively simple robot achieves autonomous phototaxis, as it steers in place when only one eye detects light and implicitly coordinates the limbs to hop forward when both eyes do (fig.~S22 and movie~S7). Outside of the lab, this allows the robot to move from a dark room to a brighter one (movie~S7) and to continuously follow an operator who carries a light (Fig.~5E and movie~S7).

\paragraph*{Conclusions}

In conclusion, we leverage kinks traveling along a soft tube to create a self-oscillating limb that cyclically performs a full-step motion at high frequencies. Inspired by nature, we blur the boundaries between actuation, control, and body-environment feedback by physically synchronizing multiple limbs via explicit and implicit couplings, realizing rapidly moving, autonomous robots. The observations from the various robotic demonstrations we performed point towards the potential of a holistic and overarching approach when designing robots, to achieve robust and adaptive behavior across diverse environments. This approach does place emphasis on the physical design of the highly interacting robot, and therefore, it will likely require the development and improvement of available design tools for these types of integrated systems to discover or design useful and robust emergent dynamic behaviors.



\section*{Acknowledgments}

We thank all members of our Soft Robotic Matter Group for the invaluable discussions. We thank Niels Commandeur for technical support, Sergio Picella for the input on the analytical model, and Corentin Coulais and Said RK Rodríguez for the constructive feedback. \textbf{Funding:} J.T.B.O. acknowledges the European Union’s 2020 ERC-STG under grant agreement No. 948132. This work is part of the Dutch Research Council (NWO) and was performed at the research institute AMOLF. \textbf{Authors contributions:} A.C. and J.T.B.O. proposed and developed the research idea; A.C. designed and fabricated the devices; A.C. and H.A.H.S. performed the experiments; A.C. and H.A.H.S. performed the data analysis; A.C. created the figures and videos; A.C. and J.T.B.O. wrote the manuscript; A.C., H.A.H.S., and J.T.B.O. revised the manuscript; J.T.B.O. supervised the research. \textbf{Competing interests:} A.C., H.A.H.S., and J.T.B.O. declare no competing interests. \textbf{Data and materials availability:} All data is available on Zenodo \cite{zenodo}.


\section*{Supplementary Materials}

\textbf{The PDF file includes:}

Materials and Methods

Supplementary Text

Figs. S1 to S27

Tables S1 to S4

Captions for Movies S1 to S7

References \textit{(43-62})\\
\textbf{Other Supplementary Material for this manuscript includes the following:}

Movies S1 to S7

\end{document}